\documentclass[a4paper,10pt]{article}
\usepackage[utf8x]{inputenc}

\title{ Gauge and Supersymmetric Invariance of 
a Boundary Bagger-Lambert-Gustavsson Theory }
\author{ Mir Faizal \\
 \\ 
Mathematical Institute, University of Oxford
\\ Oxford
OX1 3LB, United Kingdom 
 }

\begin{document}

\maketitle

\begin{abstract}
In this paper we will discuss the effect of a having a boundary on the supersymmetric invariance and gauge invariance of 
the Bagger-Lambert-Gustavsson (BLG) Theory.
We will show that even though the supersymmetry and gauge invariance of the 
original BLG theory is broken due to the presence of a boundary, it restored by the addition of  
 suitable boundary terms. In fact, to achieve the gauge invariance of this theory, 
we will have to  introduce new boundary degrees of freedom. 
The boundary theory obeyed by these new boundary degrees of freedom  will  be shown to be a generalization of the 
gauged Wess-Zumino-Witten model, with the generators of the Lie algebra replaced by the generators of the Lie $3$-algebra. 
The gauge and supersymmetry variations
 of the  boundary theory will exactly cancel the boundary 
terms generated by the  gauge and supersymmetric variations of the bulk theory.

\end{abstract}
\section{Introduction}
 Bagger-Lambert-Gustavsson (BLG) theory is a superconformal field theory with $\mathcal{N} =8$ supersymmetry
\cite{BL1, BL2, BL3, blG, blg15}. This is thought to be the 
theory that describes the low energy behavior of would volume  of multiple $M2$-branes. In this theory fields take value in a  Lie $3$-algebra rather than a regular Lie algebra. 
The BLG theory has been analysed in $\mathcal{N} =1$ superspace \cite{14,14abjm}, 
$\mathcal{N } =2$ superspace \cite{214,214a}, 
and $\mathcal{N } =8$ superspace \cite{8a1, 8a2}. 
In this paper we will perform our analysis in $\mathcal{N} =1$ superspace.
Higgs mechanism for the BLG theory has also been studied in  $\mathcal{N} =1$ superspace formalism \cite{14abjm}. Thus, 
higher derivative corrections to super-Yang-Mills on $D2$-branes  have been analysed in $\mathcal{N} =1$ superspace.

Just like in string theory, strings can end on $D$-branes, in $M$-theory $M2$-branes can end on  $M5$-branes. 
So, the $M2$-brane can be viewed  as  string like
soliton of the non-linear world volume equations of motion. In fact,  this  soliton is known as
the self dual string \cite{sds}. In analogy with $ D1$-$D3 $ system
this the ending can also be described as a fuzzy $S^3$ funnel solution of
the Basu-Harvey equation \cite{bh1, bh2, bh4}. 
Thus, by studding $M2$-branes with boundaries, we can understand the dynamics of  $M5$-branes. In fact, the equation of motion of a 
$M5$-brane have been derived by demanding the $\kappa$-symmetry of the open membrane action \cite{oma1, oma2}. 

In addition to this  $M2$-branes ending on $M9$-branes and gravitational waves have also been analysed \cite{BCIntMem}. 
 Boundary effects for the BLG theory with fluxes, on a manifold with boundaries,  have  been studied  \cite{ChuSehmbi, d12}.
The fact that the fields take values in a Lie $3$-algebra is crucial in the construction of the  the BLG theory. 
However, only one  example of a Lie $3$-algebra is known and it has been difficult to increase the rank of the gauge group. 
So, usually another theory called the ABJM theory is studied as the theory of multiple $M2$-branes \cite{ abjma}.
This theory has only manifest 
$\mathcal{N} =6$ supersymmetry, but it is expected to be enhanced to full $\mathcal{N} =8$ supersymmetry \cite{abjma1}. This 
theory agrees with the BLG theory for the only example of the Lie $3$-algebra known \cite{abjma2}.
The supersymmetric and gauge invariance of the ABJM theory in presence of boundaries has been already been studied.  
In fact, a gauge invariant model has been constructed by 
coupling the bosonic ABJM in presence of boundaries to a gauged Wess-Zumino-Witten model living on the boundary \cite{20}. 
The supersymmetric invariance of general field theories in presence of boundaries has been studied in $\mathcal{N} =1$ superspace formalism
 \cite{21}. This formalism has also been used to analyse the supersymmetry of the 
gauge part of the ABJM theory \cite{MemBdry}. 

In this paper we will analyse the BLG theory with a boundary in $\mathcal{N} =1$ superspace formalism. We will observe that like the ABJM theory, both 
 supersymmetric and gauge invariant is broken due to the presence of a boundary. However if we suitable  coupled the 
bulk BLG theory to a boundary theory, the resultant theory can be made both gauge invariant and 
supersymmetric. We will find that the theory that is needed to make the bulk BLG theory 
gauge invariant is a generalization of the gauged Wess-Zumino-Witten model, with the gauge symmetries generated by the Lie $3$-algebra.

\section{Super-Covariant Derivatives}
Three dimensional non-abelian supergauge theory in $\mathcal{N} = 1$ superspace formalism has been throughly analysed \cite{1001SUSY} and boundary 
effects in abelian  supergauge theory have also been studied \cite{21}. In this section we will analyse the boundary effects in supergauge theory, with 
the gauge symmetry generated by generators of the Lie $3$-algebra. To do so we first review some properties of a Lie $3$-algebra \cite{oneloop}. 
A Lie 3-algebra ${\cal{A}}$ is defined by 
\begin{equation}
[T^A,T^B,T^C] = f^{ABC}_D T^D.
\end{equation}
where $T^A$ are the generators of this Lie $3$-algebra and 
 $f^{ABC}_D$ are the structure constants. These structure constants are totally antisymmetric 
  in $A,B,C$ and are subject to the fundamental identity resembling the Jacobi identity for Lie algebras, 
\begin{equation}
f^{[ABC}_G f^{D]EG}_H = 0. \label{13}
\end{equation}
It is useful to define another constant constructed from these structure constant as 
\begin{equation}
C^{AB,CD}_{EF} = 2f^{AB[C}_{[E} \delta^{D]}_{F]}.
\end{equation}
These constants are anti-symmetric in the  pair
 of indices $AB$ and $CD$ and also satisfy a kind of Jacobi 
identity \cite{oneloop},
\begin{equation}
C^{AB,CD}_{EF} C^{GH,EF}_{KL} + C^{GH,AB}_{EF} C^{CD,EF}_{KL} 
+C^{CD,GH}_{EF} C^{AB,GH}_{KL} =0.\label{13a}
\end{equation}
This Lie $3$-algebra is also accompanied by a metric $h^{AB} = Tr (T^A T^B)$ which is used to raise and lower the indices. 

Now the scalar superfields   $X^I$ and $X^{\dagger}$ are suitably
 contracted with the generators of this Lie $3$-algebra, 
 $X^I= X^{A I} T_A,$ and $ 
X^{I \dagger}= X^{I A \dagger} T_A$,
and  transform under infinitesimal gauge transformations as
\begin{eqnarray}
  \delta X^{I A}&=&  i\Lambda^{AB} X^I_B,\nonumber\\
\delta X^{I A \dagger} &=& -i X^{I \dagger}_B\Lambda^{AB}.
\end{eqnarray}
It is convenient to define $\Lambda = \Lambda^{AB} T_A T_B$ and the following super-covariant derivatives 
of these superfields
\begin{eqnarray}
  \nabla_a  X^I_A&=& D_a X_A -i f^{BCD}_A \Gamma_{a BC} X^{I}_{ D},\nonumber\\
 \nabla_a X^{I \dagger}_A &=& D_a X^{I \dagger}_A + if^{BCD}_A X^{I \dagger}_D \Gamma^{a BC},
\end{eqnarray}
where
 \begin{equation}
 D_a = \partial_a + (\gamma^\mu \partial_\mu)^b_a \theta_b.
\end{equation}
 It is also convenient to define  $\Gamma_a$ as a matrix valued spinor
 superfield suitable contracted with generators of a Lie $3$-algebra,  
$\Gamma_a = \Gamma_a^{AB} T_AT_B $.
This matrix valued spinor superfield  transforms under 
gauge transformations as 
\begin{equation}
 \delta \Gamma_a = \nabla_a \Lambda,
\end{equation}
where 
\begin{equation}
  \nabla_a \Lambda = [D_a \Lambda_{AB} + C^{CD,EF}_{AB}\Gamma_{CD a} \Lambda_{EF}]T^A T^B.
\end{equation}
In fact, we will define the covariant derivative on $\Gamma_a^{AB}$ as follows 
\begin{equation}
 ( \nabla_a \Gamma_b)_{AB} = D_a \Gamma_{b AB} + C^{CD,EF}_{AB}\Gamma_{CD a} \Gamma_{b EF}.
\end{equation}
It is useful to define ordinary gauge covariant derivatives for component of matter and gauge superfields. Thus, if $a^A_i$ and $\overline{a}^A_i$ are 
components of the matter fields then the ordinary gauge covariant derivatives for them are given by 
\begin{eqnarray}
   \mathcal{D}_\mu a_i^A &=& 
\partial_\mu a_i^A -i f^{ABC}_D A_{\mu BC} a_{i }^D, \nonumber \\ 
  \mathcal{D}_\mu \overline{a}_i^A &=& 
\partial_\mu \overline{a}_i^A + if^{ABC}_D \overline{a}_{i}^D A_{ \mu BC}.
\end{eqnarray}
Similarly, the ordinary gauge covariant derivative for the component fields $e^{AB}_i$ of the gauge superfield is given by 
\begin{equation}
 \mathcal{D}_\mu e_{i AB} = 
\partial_\mu e_{i AB} + C^{CD,EF}_{AB} A_{\mu CD } e_{  i EF }.
\end{equation}
 Here $A_{\mu}^{AB}$ is the ordinary gauge connection which is given by 
\begin{equation}
 A^{\mu AB} = - \frac{1 }{2} [ (\nabla^a (\gamma^{\mu })_a^b \Gamma_b)^{AB} ]_|,
\end{equation}
where $'|'$  means that the quantity is evaluated at $\theta_a =0$. 

Now we can calculate the following 
\begin{eqnarray}
 (\{\nabla_a , \nabla_b\} X^I)_{A} &=& ( \nabla_a \nabla_b X^I)_A + ( \nabla_b \nabla_a X^I)_A \nonumber \\ 
&=& (D_a \delta^D_A -  i f^{BCD}_A \Gamma_{a CD})\nonumber \\ && \times( D_b \delta^G_D - i f^{EFG}_D \Gamma_{b EF}) X_G^I \nonumber \\ 
&& - (D_b \delta^D_A -  i f^{BCD}_A \Gamma_{b CD})\nonumber \\ && \times( D_a \delta^G_D - i f^{EFG}_D \Gamma_{a EF}) X_G^I \nonumber \\ 
&=& 2 (\gamma_{ab}^\mu \mathcal{D}_\mu X^I)_A,
\end{eqnarray}
and 
\begin{eqnarray}
 (\{  \nabla_a ,   \nabla_b\} X^{I\dagger})_{A} &=& (  \nabla_a  \nabla_b X^I)_A + 
(  \nabla_b  \nabla_a X^{I\dagger})_A \nonumber \\ 
&=& (D_a \delta^D_A +  i f^{BCD}_A \Gamma_{a CD})\nonumber \\ && \times( D_b \delta^G_D + i f^{EFG}_D \Gamma_{b EF}) X_G^{I\dagger} \nonumber \\ 
&& - (D_b \delta^D_A +  i f^{BCD}_A \Gamma_{b CD})\nonumber \\ && \times( D_a \delta^G_D + i f^{EFG}_D \Gamma_{a EF}) X_G^{I \dagger} \nonumber \\ 
&=& 2 (\gamma_{ab}^\mu \mathcal{D}_\mu X^{I\dagger})_A.
\end{eqnarray}
We also have 
\begin{eqnarray}
 (\{ \nabla_a ,  \nabla_b\} \Gamma_c)_{AB} &=& ( \nabla_a  \nabla_b \Gamma_c )_{AB} + 
(  \nabla_b  \nabla_a \Gamma_c)_{AB} \nonumber \\ 
&=& (D_a \delta^A_E \delta^E_F +  C^{CD, EF}_{AB} \Gamma_{a CD})
 \nonumber \\ && \times( D_b \delta^L_E \delta^M_F +  C^{GH, LM}_{EF} \Gamma_{b GH})\Gamma_{c LM} \nonumber \\ 
&& - (D_b \delta^A_E \delta^E_F +  C^{CD, EF}_{AB} \Gamma_{b CD})\nonumber \\ && \times( D_a \delta^L_E \delta^M_F +  C^{GH, LM}_{EF} \Gamma_{a GH})\Gamma_{c LM} \nonumber \\ 
&=& 2 (\gamma_{ab}^\mu \mathcal{D}_\mu \Gamma_c)_{AB}.
\end{eqnarray}
Using $X^I_A T^A = X^I$, $X^{I\dagger}_A T^A = X^{I\dagger}$ and $\Gamma_{c AB} T^A T^B = \Gamma_c$, we have 
\begin{eqnarray}
  \{ \nabla_a , \nabla_b\} X^I &=& 2 \gamma_{ab}^\mu \mathcal{D}_\mu X^I, \nonumber \\ 
  \{\nabla_a , \nabla_b\} X^{I \dagger} &=& 2 \gamma_{ab}^\mu \mathcal{D}_\mu X^{I \dagger}, \nonumber \\ 
  \{\nabla_a , \nabla_b\} \Gamma_c &=& 2 \gamma_{ab}^\mu \mathcal{D}_\mu \Gamma_c.
\end{eqnarray}
Now we can write $(\nabla_a \nabla_b X^I)_A, (\nabla_a \nabla_bX^{I \dagger})_A$ and $(\nabla_a \nabla_b \Gamma_c)_{AB}$
 as a half of the sum of there commutator with its anticommutator, 
\begin{eqnarray}
 ( \nabla_a \nabla_b X^I )_A&=& \frac{1}{2} (\{ \nabla_a, \nabla_b\}X^I)_A + \frac{1}{2}
 ([\nabla_a, \nabla_b]X^I)_A, 
\nonumber \\ 
 ( \nabla_a \nabla_b X^{I \dagger} )_A&=&  \frac{1}{2} (\{ \nabla_a, \nabla_b\}X^{I \dagger} )_A + \frac{1}{2}
 ([\nabla_a, \nabla_b]X^{I \dagger} )_A, 
\nonumber \\ 
 (  \nabla_a  \nabla_b \Gamma_c )_{AB} &=& \frac{1}{2} (\{  \nabla_a,  \nabla_b\}\Gamma_c)_{AB} + \frac{1}{2}
 ([ \nabla_a,  \nabla_b]\Gamma_c)_{AB}.
\end{eqnarray}
However, for  $\mathcal{N} =1$ superfields 
 in three dimensions the indices $'a'$ are two-dimensional. Thus, 
the anticommutator   $([\nabla_a , \nabla_b] X^I)_A $, $ ([\nabla_a , \nabla_b] X^{I \dagger})_A $, and $ ([\nabla_a , \nabla_b]\Gamma_c)_{AB}$
 must be proportional
 to  anti-symmetric tensors some $(C_{ab}X^I)_A $, $  (C_{ab}X^{I \dagger})_A $, and $  (C_{ab}\Gamma_c)_{AB}$, respectively. So, we find 
\begin{eqnarray}
(\nabla_a \nabla_b X^I)_A = ((\gamma^\mu _{ab} \mathcal{D}_\mu - C_{ab} \nabla^2 ) X^I)_A, \nonumber \\ 
(\nabla_a \nabla_b X^{I \dagger})_A = ((\gamma^\mu _{ab} \mathcal{D}_\mu -  C_{ab}  \nabla^2) X^{I \dagger})_A, \nonumber \\ 
 (\nabla_a \nabla_b \Gamma_{c })_{AB} = ((\gamma^\mu _{ab} \mathcal{D}_\mu -  C_{ab} \nabla^2 )\Gamma_c)_{AB}.
\end{eqnarray}
Now using \cite{21}
\begin{equation}
  D_a D_b D_c = \frac{1}{2} D_a \{D_b, D_c\} -\frac{1}{2}
  D_b \{D_a, D_c \}
 + \frac{1}{2}D_c \{D_a ,D_b \}, 
\end{equation}
along with Eqs. (\ref{13}) and (\ref{13a}), we get, 
\begin{eqnarray}
 ( \nabla_a \nabla_b \nabla_c X^I)_A&=& \frac{1}{2} 
(\nabla_a \{\nabla_b, \nabla_c\} X^I)_A-  \frac{1}{2}
(\nabla_b \{\nabla_a, \nabla_c \}X^I)_A
 \nonumber \\ &&+ \frac{1}{2}( \nabla_c \{\nabla_a ,\nabla_b \}X^I)_A ,
\nonumber \\ 
  (\nabla_a \nabla_b \nabla_cX^{I \dagger})_A &=&
 \frac{1}{2}( \nabla_a \{\nabla_b, \nabla_c\} X^{I \dagger})_A-  \frac{1}{2}
(\nabla_b \{\nabla_a, \nabla_c \}X^{I \dagger})_A
 \nonumber \\ &&+ \frac{1}{2} (\nabla_c \{\nabla_a ,\nabla_b \}X^{I \dagger})_A,\nonumber \\ 
 ( \nabla_a \nabla_b \nabla_c\Gamma_{d })_{AB}  &=& 
\frac{1}{2}( \nabla_a \{\nabla_b, \nabla_c\}\Gamma_{d })_{AB} 
 -  \frac{1}{2}(\nabla_b \{\nabla_a, \nabla_c \}\Gamma_{d })_{AB} 
 \nonumber \\ &&+ \frac{1}{2}
( \nabla_c \{\nabla_a ,\nabla_b \}\Gamma_{d })_{AB} .
\end{eqnarray}
Thus, we get 
\begin{equation}
 (\nabla^a \nabla_b \nabla_aX^I)_A =
(\nabla^a \nabla_b \nabla_a X^{I \dagger})_A=((\nabla^a \nabla_b \nabla_a \Gamma_{c })_{AB}=0,
\end{equation}
and 
\begin{eqnarray}
(\nabla^2 \nabla_a  X^I)_A = ((\gamma^\mu \nabla )_{a} \mathcal{D}_\mu X^I)_A , \nonumber \\ 
(\nabla^2 \nabla_a  X^{I \dagger})_A = ((\gamma^\mu \nabla )_{a} \mathcal{D}_\mu X^{I \dagger})_A, \nonumber \\ 
(\nabla^2 \nabla_a \Gamma_{c })_{AB} = ((\gamma^\mu \nabla )_{a} \mathcal{D}_\mu\Gamma_c)_{AB}. \label{22}
\end{eqnarray}

\section{Boundary Super-Covariant Derivatives}
In the previous section we analysed some properties of super-covariant derivatives for a gauge theory where the gauge symmetry is generated 
by a Lie $3$-algebra. In this section we will analyse the effect of having a boundary on these super-covariant derivatives. We start by having 
 a boundary at fixed $x^3$. Thus our indices for the coordinates 
 will splits as  $\mu =(m, 3 )$. We will denoted the induced value of the fields $X, X^{\dagger},\Gamma_a, \Lambda$ on the boundary 
 by $X', {X^{\dagger}}',\Gamma_a', \Lambda'$ and the 
 induced value of the super-derivative $D_a$ and the 
super-covariant derivative $\nabla_a$ on the boundary will be denoted by 
$D_a'$ and $\nabla_a'$, respectively.  
This boundary super-derivative $D_a'$ is obtained by neglecting 
$\gamma^3 \partial_3$ contributions in $D_a$, 
 \begin{equation}
 D_a' = \partial_a + (\gamma^m \partial_m)^b_a \theta_b.
\end{equation}
The boundary super-covariant derivative $\nabla_a'$ can thus be written as  
\begin{eqnarray}
  (\nabla_a'  {X^I}')_A &=& D_a' {X^I}'_A -if^{BCD}_A\Gamma_{BC a }'{X^{I}}'_D,\nonumber\\
(\nabla_a'{X^{I \dagger}}')_A &=& D_a'{X^{I \dagger}}'_A
 + i f^{BCD}_A {X^{\dagger}}'_D\Gamma_{BC a }', \nonumber \\ 
  (\nabla_a' \Gamma_b')_{AB} &=& D_a' \Gamma_{b AB}' + C^{CD,EF}_{AB}
\Gamma_{CD a}' \Gamma_{b EF}', \nonumber \\ 
 (\nabla_a' \Lambda')_{AB} &=& 
D_a' \Lambda'_{AB} + C^{CD,EF}_{AB}\Gamma'_{CD a} \Lambda'_{EF}.
\end{eqnarray}
In fact, this notation will be used to denote all  boundary quantities along with the 
induced value of any bulk quantity on the boundary. 
Thus, this  convention will be followed even for component fields of superfields.  
Now we define projection operators $P_{\pm}$ as
\begin{equation}
 (P_{\pm  })_{ab}  = \frac{1}{2} (C_{ab} \pm (\gamma^{3})_{ab}). \label{a}
\end{equation}
These projection operators can be used to project the super-covariant derivative $\nabla_a$ as, 
$
 \nabla_{ \pm b }= (P_{\pm})^a_b \nabla_a, 
$
and  $\nabla'_{\pm b}$ as, 
$
 \nabla'_{\pm b } = (P_{\pm })^a_b \nabla'_a, 
$
where $\nabla_{\pm a}'$ is the induced value of $\nabla_{\pm a}$ on the boundary. 
Similarly, for the ordinary gauge covariant derivatives we have 
$
(P_{\pm} \gamma^m)_{ab} \mathcal{D}_m = (\gamma^{\pm})_{ab} \mathcal{D}_{\pm}
$,
where $\gamma^{\pm} = \gamma^0 \pm \gamma^1$ and
$\mathcal{D}_{\pm} = \frac{1}{2} (\mathcal{D}_0 \pm \mathcal{D}_1)$.
Now we can calculate the projected values of the super-covariant derivative acting on $X^I_A$ 
\begin{eqnarray}
 (\nabla_{+a} \nabla_{+b} X^I)_A&=&((P_+)_a^c (P_+)_b^d
 (\nabla_{c} \nabla_{d} X^I)_A\nonumber \\ &=&
  - (( \gamma^+)_{ab} \mathcal{D}_+X^I)_A, \\
 (\nabla_{-a} \nabla_{-b}X^I)_A &=& ((P_-)_a^c (P_-)_b^d 
(\nabla_{c} \nabla_{d} X^I)_A\nonumber \\ &=&
 - ((\gamma^-)_{ab} \mathcal{D}_-X^I)_A, \\
( \nabla_{-a} \nabla_{+b}X^I)_A &=&
 ((P_-)_a^c (P_+)_b^d (\nabla_{c} \nabla_{d} X^I)_A\nonumber \\ 
&=&  -((P_-)_{ab} (\mathcal{D}_3 + \nabla^2)X^I)_A,\\  
 (\nabla_{+a} \nabla_{-b}X^I)_A &=&
 ((P_+)_a^c (P_-)_b^d (\nabla_{c} \nabla_{d} X^I)_A\nonumber \\ &=&  
-((P_+)_{ab} (-\mathcal{D}_3 +\nabla^2 )X^I)_A. 
 \end{eqnarray}
Similarly, we can calculate the projected values of the super-covariant derivative acting on $X^{I \dagger }_A$
\begin{eqnarray}
 (\nabla_{+a} \nabla_{+b} X^{I \dagger })_A&=& ((P_+)_a^c (P_+)_b^d (\nabla_{c} \nabla_{d}X^{I \dagger })_A\nonumber \\ &=& - (( \gamma^+)_{ab} \mathcal{D}_+X^{I \dagger })_A, \\
 (\nabla_{-a} \nabla_{-b}X^{I \dagger })_A &=& ((P_-)_a^c (P_-)_b^d (\nabla_{c} \nabla_{d} X^{I \dagger })_A\nonumber \\ &=& - ((\gamma^-)_{ab} \mathcal{D}_-X^{I \dagger })_A, \\
( \nabla_{-a} \nabla_{+b}X^{I \dagger })_A &=& ((P_-)_a^c (P_+)_b^d  (\nabla_{c} \nabla_{d} X^{I \dagger })_A\nonumber \\ &=&  -((P_-)_{ab} (\mathcal{D}_3 + \nabla^2)X^{I \dagger })_A,\\  
 (\nabla_{+a} \nabla_{-b}X^{I \dagger })_A &=& ((P_+)_a^c (P_-)_b^d (\nabla_{c} \nabla_{d} X^{I \dagger })_A\nonumber \\ &=& - ((P_+)_{ab} (-\mathcal{D}_3 +\nabla^2 )X^{I \dagger })_A. 
 \end{eqnarray}
Finally we can calculate the projected value of the super-covariant derivative acting on $\Gamma_c^{AB}$
\begin{eqnarray}
 (\nabla_{+a} \nabla_{+b} \Gamma_c)_{AB}&=& ((P_+)_a^c (P_+)_b^d (\nabla_{c} \nabla_{d} \Gamma_c)_A\nonumber \\ &=& - (( \gamma^+)_{ab} \mathcal{D}_+X^I)_{AB}, \\
 (\nabla_{-a} \nabla_{-b}\Gamma_c)_{AB} &=&((P_-)_a^c (P_-)_b^d(\nabla_{c} \nabla_{d} \Gamma_c)_A\nonumber \\ &=&  - (( \gamma^-)_{ab} \mathcal{D}_-X^I)_{AB}, \\
( \nabla_{-a} \nabla_{+b}\Gamma_c)_{AB}&=& ((P_-)_a^c (P_+)_b^d (\nabla_{c} \nabla_{d} \Gamma_c)_A\nonumber \\ &=&  -((P_-)_{ab} (\mathcal{D}_3 + \nabla^2)X^I)_{AB},\\  
 (\nabla_{+a} \nabla_{-b}\Gamma_c)_{AB} &=& ((P_+)_a^c (P_-)_b^d (\nabla_{c} \nabla_{d} \Gamma_c)_A\nonumber \\ &=&  -((P_+)_{ab} (-\mathcal{D}_3 +\nabla^2 )X^I)_{AB}. 
 \end{eqnarray}
Suitably contracting the fields with generators of the Lie $3$ algebra as $X^I_A T^A = X^I$, $X^{I\dagger}_A T^A = X^{I\dagger}$ 
and $\Gamma_{c AB} T^A T^B = \Gamma_c$, 
we get the following algebra 
\begin{eqnarray}
 \{ \nabla_{ + a}, \nabla_{ +b} \} X^I= - 2  (\gamma^+)_{ab} \mathcal{D}_+  X^I, &
  \{ \nabla_{-a }, \nabla_{-b} \} X^I=-2  ( \gamma^-)_{ab} \mathcal{D}_-X^I, &\nonumber \\
 \{ \nabla_{- a}, \nabla_{ +b} \}X^I = -2  (P_-)_{ab} \mathcal{D}_3 X^I, & 
 \{ \nabla_{ + a}, \nabla_{ +b} \} X^{I\dagger}= - 2  (\gamma^+)_{ab} \mathcal{D}_+X^{I\dagger},& \nonumber \\
  \{ \nabla_{-a }, \nabla_{-b} \} X^{I\dagger}= - 2  ( \gamma^-)_{ab} \mathcal{D}_-X^{I\dagger}, &
 \{ \nabla_{- a}, \nabla_{ +b} \}X^{I\dagger} = -2  (P_-)_{ab} \mathcal{D}_3X^{I\dagger},&\nonumber \\ 
 \{ \nabla_{ + a}, \nabla_{ +b} \} \Gamma_c= - 2  (\gamma^+)_{ab} \mathcal{D}_+\Gamma_c, &
  \{ \nabla_{-a }, \nabla_{-b} \} \Gamma_c= - 2  ( \gamma^-)_{ab} \mathcal{D}_-\Gamma_c,& \nonumber \\
 \{ \nabla_{- a}, \nabla_{ +b} \}\Gamma_c =  -2  (P_-)_{ab} \mathcal{D}_3\Gamma_c.& 
\end{eqnarray}
Now we have effectively decomposed $\mathcal{N} =1 $ superfields in three dimensions into $\mathcal{N} = (1,0)$ in two dimensions. The 
generators of the $\mathcal{N} =1$ supersymmetry in three dimensions is 
\begin{equation}
 Q_a   = \partial_a - (\gamma^\mu \partial_\mu)^b_a \theta_b.
\end{equation}
This generator of $\mathcal{N} =1$ supersymmetry in three dimensions splits into $Q_{\pm a} = P_{\pm a}^b Q_b$ in two dimensions, 
\begin{equation}
  \epsilon^a Q_a = \epsilon^{a -} Q_{- a} + \epsilon^{ a + } Q_{+ a},
\end{equation}
where $\epsilon_{a \pm} = P_{\pm a }^b \epsilon_b $. It can be shown that these generators of the supersymmetry in presence of a boundary satisfy 
\begin{eqnarray}
 \{ Q_{ + a}, Q_{ +b} \} = 2  (\gamma^+)_{ab} \partial_+, &&
  \{ Q_{-a }, Q_{-b} \} = 2  ( \gamma^-)_{ab} \partial_-, \nonumber \\
 \{ Q_{- a}, Q_{ +b} \} = 2  (P_-)_{ab} \partial_3,
\end{eqnarray}
where $\partial_{\pm} = \frac{1}{2} (\partial_0 \pm \partial_1)$. It may be noted in absence of a boundary term this algebra is actually the algebra of
two copies of $\mathcal{N} =1$ supersymmetric theories in two dimensions.  

\section{Supersymmetry in Presence of a Boundary }
To analyse the boundary effects in the BLG theory we first construct 
the BLG theory on manifolds without boundaries. 
The Lagrangian  for the BLG theory in $\mathcal{N} =1 $ superspace formalism is given by 
\begin{equation}
 \mathcal{L}_{BLG} = \mathcal{L}_{CS}  + \mathcal{L}_{KE} +  \mathcal{L}_{V},
\end{equation}
 where $\mathcal{L}_{CS}$ is the Chern-Simon term, $ \mathcal{L}_{KE}$ is the the kinetic energy term
and $\mathcal{L}_{V}$ is the potential energy part term of the BLG theory. 
In our notations a trace over the Lie $3$-algebra will be assumed in the Lagrangian of the BLG theory. 
Now the Chern-Simon term is given by 
\begin{equation}
 \mathcal{L}_{CS} = -\frac{k}{4\pi}\nabla^2   [f^{ABCD}\Gamma^{a}_{ AB}  \Omega_{a CD} ]_|.
\end{equation}
 where
\begin{eqnarray}
 \Omega_{ a AB} & = & \omega_{a AB} - \frac{1}{3}C^{CD,EF}_{AB}[\Gamma^{b CD}, \Gamma_{ab EF}] \\
 \omega_{a AB} & = & \frac{1}{2} D^b D_a \Gamma_{b AB} 
-i  C^{CD,EF}_{AB}[\Gamma^b_{CD} , D_b \Gamma_{a EF}] \nonumber \\ && -
 \frac{1}{3}C^{CD,EF}_{AB} C^{GH,IJ}_{EF}[ \Gamma^b_{CD} ,
\{ \Gamma_{b GH} , \Gamma_{a IJ}\} ], \label{omega} \\
 \Gamma_{ab AB} & = & -\frac{i}{2} \left[ D_{(a}\Gamma_{b) AB} 
- 2 i C^{CD,EF}_{AB}\{\Gamma_{a CD}, \Gamma_{b EF}\} \right] .
\end{eqnarray}  
The covariant divergence of $\omega_{a AB}$ vanishes
\begin{eqnarray}
( \nabla^a \omega_a)_{AB} &=&[D^a \delta^E_A \delta^F_B + C_{AB}^{CD, EF}\Gamma^a_{CD}] \omega_{a EF} \nonumber \\ &=&  
  \frac{1}{2} \delta^E_A \delta^F_B D^a D^b D_a \Gamma_{b EF} -i  C^{CD,LM}_{EF}\delta^E_A \delta^F_B D^a[\Gamma^b_{CD} , D_b \Gamma_{a LM}] 
\nonumber \\ && -
 \frac{1}{3}C^{CD,LM}_{EF} C^{GH,IJ}_{LM} \delta^E_A \delta^F_B D^a[ \Gamma^b_{CD} ,
\{ \Gamma_{b GH} , \Gamma_{a IJ}\} ] 
\nonumber \\ &&
 -i  C_{AB}^{CD, EF}C^{IJ,LM}_{EF}\Gamma^a_{CD}[\Gamma^b_{IJ} , D_b \Gamma_{a LM}] 
\nonumber \\ &&  -
 \frac{1}{3}C^{CD,LM}_{EF} C^{GH,IJ}_{LM}C_{AB}^{ST, EF}
\Gamma^a_{CD}[ \Gamma^b_{ST} ,
\{ \Gamma_{b GH} , \Gamma_{a IJ}\} ] \nonumber \\ && + 
\frac{1}{2} C_{AB}^{CD, EF}\Gamma^a_{CD} D^b D_a \Gamma_{b EF} 
 \nonumber \\ &=& 
 0, \label{cov}
\end{eqnarray}
here we have used  $D^a D_b D_a =0$, \cite{21}. 
Now the components of $\Gamma_{a}^{AB}$ are given by 
\begin{eqnarray}
 \chi^{AB}_a = [\Gamma^{AB}_a]_|, && A^{AB} = - \frac{1}{2}[(\nabla^a \Gamma_a)^{AB}]_|, \nonumber \\ 
A^{\mu AB} = - \frac{1 }{2} [ (\nabla^a (\gamma^{\mu })_a^b \Gamma_b )^{AB}]_|, &&
 E^{AB}_a = - [(\nabla^b \nabla_a  \Gamma_b)^{AB}]_|, 
\end{eqnarray}
and so the components of the $\omega^{AB}_a$  can now be written as   
\begin{eqnarray}
 [ (\nabla^a (\gamma^{\mu })_a^b \omega_b)^{AB} ]_|=  \epsilon^{\mu \nu \rho} F^{AB}_{\nu \rho},  &&
[(\nabla^a \omega_a)^{AB}]_| =0,
 \nonumber \\ 
 - [(\nabla^b \nabla_a  \omega_b)^{AB}]_| = 2( (\gamma^\mu \mathcal{D}_\mu)_a^b E_b)^{AB}, &&[\omega^{AB}_a]_| =  E^{AB}_a , 
\end{eqnarray}
where $\epsilon_{\mu \nu \rho} $ is an anti-symmetric tensor. Thus, 
 the component form for  Chern-Simons 
term  can be written as
\begin{eqnarray}
 \mathcal{L}_{CS} &=& \frac{k}{4\pi}\left[ \epsilon^{\mu \nu \rho} \left(f^{ABCD} A_{ \mu AB} \partial_\nu  A^{ \rho CD} 
 + \frac{i}{3}C_{AB}^{CD, EF} A_{\mu }^{AB} A_{\nu CD}  A_{\rho EF}  \right. \right)  \nonumber \\  &&\left.  + 
f^{ABCD}( E^{a }_{AB} E_{a CD} + (\mathcal{D}_\mu  
\chi^a(\gamma^\mu)_a^b)_{AB} E_{b CD} 
\right. \nonumber \\  &&\left.+   (\chi^a(\gamma^\mu)_a^b)_{AB} (\mathcal{D}_\mu E_b)_{CD} )\right].
\end{eqnarray}
The Kinetic energy term for the BLG theory can be written as  
\begin{eqnarray}
 \mathcal{L}_{KE}
&=&-\frac{1}{4}  \nabla^2 [ (\nabla^a X^I)^A  (\nabla_a X^{\dagger}_I)_A ]_|,
\end{eqnarray}
and the potential term for the BLG theory can be written as  
\begin{eqnarray}
 \mathcal{L}_{V} &=& 
-\frac{2\pi}{k}\nabla^2 
[\epsilon_{IJKL} f^{ABCD}X^I_A X^{K \dagger}_B X^{J}_C  X^{L \dagger}_D]_|.
\end{eqnarray}
This theory is manifestly invariant under $\mathcal{N} =1$ superspace transformations, even thought in reality it has $\mathcal{N} = 8$ supersymmetry.
Thus, we have 
\begin{eqnarray}
 \delta_{S}  \mathcal{L}_{BLG} &=&   \epsilon^a Q_a  \mathcal{L}_{BLG} \nonumber \\ 
&=&  \mathcal{D}_{\mu} [\Phi (\gamma^\mu, \Gamma, X^I, X^{I \dagger})]_|. 
\end{eqnarray}
Now in absence of a boundary, we have 
\begin{equation}
  \delta_{S}  \mathcal{L}_{BLG} = 0.
\end{equation}
In fact, the supersymmetric variation of any superfield theory written in the $\mathcal{N} =1$ superspace formalism  gives 
rise to a total derivative term
\cite{21}. Thus, in absence of a boundary this term  vanishes and the theory has manifest $\mathcal{N} =1$ supersymmetry. 

Now if the full finite gauge transformation  of the superfield $\Gamma_a = \Gamma_a^{AB}T_A T_B$ is given by    
\begin{equation}
   \Gamma_{a } \rightarrow i u \, \nabla_a u^{-1},
\end{equation}
where 
\begin{equation}
 u = \exp ( i \Lambda^{AB} T_A T_B), 
\end{equation}
then the gauge transformation of the superfield $ \omega_a = \omega_a^{AB}T_A T_B $ will be given by  
\begin{equation}
  \omega_a \rightarrow u \, \omega_a u^{-1}. 
\end{equation}
Similarly, the finite gauge transformations of $X^I$ and $X^{\dagger I}$ will be given by 
\begin{eqnarray}
X^{I A}&\rightarrow&  uX^I,\nonumber\\
 X^{I A \dagger} &\rightarrow& X^{I \dagger}_Bu^{-1}.
\end{eqnarray}
Thus, under infinitesimal gauge transformations the Lagrangian
for the $\mathcal{N}=1$ non-Abelian Chern-Simons theory will
transforms as  
\begin{eqnarray}
  \delta \mathcal{L}_{CS} &=& - 
\frac{k}{4\pi} \nabla^2 [f^{ABCD} D^a \Lambda_{AB}\omega_{a CD} +
 C^{CD,EF}_{AB}\Gamma_{CD }^a 
\Lambda_{EF} \omega_{a }^{AB}]_|
\nonumber \\ &=& 
  - \frac{k}{4\pi} \nabla^2 [f^{ABCD}(\nabla^a  \Lambda )_{AB}\omega_{a CD} ]_|.
\end{eqnarray}
As the covariant derivative of $\omega_{a AB}$ vanishes, Eq. (\ref{cov}),  we get 
\begin{equation}
  \delta \mathcal{L}_{CS} =
  - \frac{k}{4\pi} \nabla^2 \nabla^a [f^{ABCD}\Lambda_{AB} \omega_{a CD} ]_|, 
\end{equation}
Now using Eq. \ref{22}, we get 
\begin{equation}
 \delta \mathcal{L}_{CS}=
 - \frac{k}{4\pi} (\gamma^\mu  \mathcal{D}_\mu \nabla)^a[ f^{ABCD}\Lambda_{AB} \omega_{a CD} ]_|.
\end{equation}
This is a total derivative, so on a manifold without a boundary, we have 
\begin{equation}
 \delta \mathcal{L}_{CS} =0. 
\end{equation}
Thus, the $\mathcal{N}=1$ non-Abelian Chern-Simons theory is invariant under these gauge 
  transformations on a manifold without a boundary. 

Now after analysing the BLG theory on manifolds without boundaries, we will analyse the effect of the boundary on the manifest $\mathcal{N} =1$ 
supersymmetric of the theory. 
As the supersymmetric variation of the BLG theory  gives rise to a total derivative term,  we have 
\begin{eqnarray}
 \delta_{S}  \mathcal{L}_{BLG} &=& 
  \epsilon^a Q_a  \mathcal{L}_{BLG}  \nonumber \\ &=& 
\mathcal{D}_{\mu} [\Phi ( \gamma^\mu, \Gamma, X^I, X^{I \dagger}) ]_|\nonumber \\ 
&\sim&   \mathcal{D}_{3} [\Phi (\gamma^3, \Gamma, X^I, X^{I \dagger})]_|. 
\end{eqnarray}
  where $\sim$ indicates 
that we have neglected the total derivative contribution along  directions other than $x^3$, as they will not contribute. So, in 
presence of a boundary, this supersymmetric variation of the BLG theory will gives rise to a boundary term,
\begin{equation}
  \delta_{S}  \mathcal{L}_{BLG} = [\Phi' (\gamma^3, \Gamma', {X^I}', {X^{I \dagger}}')]_|.
\end{equation}
Thus, the manifest $\mathcal{N} =1$ supersymmetry will be broken in presence of a boundary. 

The supersymmetry of the theory can be restored by adding a boundary term whose supersymmetric variation 
exactly   cancels the supersymmetric  variation 
of the original theory.
This has already been  done  for Abelian Chern-Simons theories 
\cite{21, MemBdry}. These results can be generalised to the present case. The term that needs to be added to the original BLG theory 
 can be written as 
\begin{equation}
 \mathcal{L}_{b,BLG} = \mathcal{L}_{b, CS}  + \mathcal{L}_{b, KE} +  \mathcal{L}_{b, V},
\end{equation}
where 
$\mathcal{L}_{b, CS}$ is the boundary term corresponding to the Chern-Simon term, $ \mathcal{L}_{b, KE}$ is the  boundary term corresponding to the
 kinetic energy term
and $\mathcal{L}_{b, V}$ is the  boundary term corresponding to the potential energy part term.
Now $\mathcal{L}_{b, CS}$ is given by 
\begin{equation}
 \mathcal{L}_{b, CS} = \frac{k}{4\pi}\mathcal{D}_3   [f^{ABCD} \Gamma^{a }_{AB}  \Omega_{a CD} ]_|,
\end{equation}
$ \mathcal{L}_{b, KE}$ is given by 
\begin{eqnarray}
 \mathcal{L}_{b, KE}
=  \frac{1}{4} \mathcal{D}_3  [ (\nabla^a X^I)^A  (\nabla_a X^{\dagger}_I)_A ]_|,
\end{eqnarray}
and $ \mathcal{L}_{b, V}$ is given by  
\begin{eqnarray}
 \mathcal{L}_{b, V} =
\frac{2\pi}{k}\mathcal{D}_3 
[\epsilon_{IJKL} f^{ABCD}X^I_A X^{K \dagger}_B X^{J}_C  X^{L \dagger}_D]_|.
\end{eqnarray}
The  resultant theory is given by the sum of these boundary terms with the  
 BLG theory,
\begin{eqnarray}
   \mathcal{L}_{s, BLG} &=& \mathcal{L}_{BLG} +   \mathcal{L}_{b, BLG} \nonumber \\ &=& 
   \frac{k}{4\pi} (-\nabla^2 + \mathcal{D}_3) \left[\frac{k}{4\pi}(f^{ABCD}\Gamma^{a }_{AB} \Omega_{a CD} +  (\frac{1}{4}\nabla^a X^I)^{AB}(
  \nabla_a X^{\dagger}_I)_{AB} \right.
\nonumber \\  && \left.+ \frac{2\pi}{k}\epsilon_{IJKL} f^{ABCD}X^I_A X^{K \dagger}_B X^{J}_C  X^{L \dagger}_D \right]_|.
\end{eqnarray}
It may be noted that only half of the  supersymmetry  of the original theory
 is preserved on the boundary. 
In this paper we will keep the 
supersymmetry generated by  to $Q_-$ and break the supersymmetry  corresponding to
$Q_+$ on the boundary,
\begin{eqnarray}
 \delta_{S}^-  \mathcal{L}_{s, BLG}&=& \epsilon^{a -} Q_{- a}  \mathcal{L}_{s, BLG}  \nonumber \\&=& 
  \epsilon^{a -} Q_{- a} \mathcal{L}_{BLG} + \epsilon^{a -} Q_{- a} \mathcal{L}_{b, BLG}\nonumber \\ 
&=& [\Phi'_- (\gamma^3, \Gamma', {X^I}', {X^{I \dagger}}') ]_|
- [\Phi'_- (\gamma^3, \Gamma', {X^I}', {X^{I \dagger}}')]_|\nonumber \\ &=&0. 
\end{eqnarray}
Thus, the supersymmetric variation of this boundary term exactly 
cancels the supersymmetric variation of the bulk Lagrangian, so 
 the sum of the bulk Lagrangian and this boundary term  preserves half of the supersymmetry of the original theory. 
It was possible to preserve the supersymmetry corresponding to $Q_+$ in presence of the boundary if the supersymmetry corresponding to 
$Q_-$ was broken. Thus, we are able to preserve either $\mathcal{N} = (1,0)$ supersymmetry or $\mathcal{N} = (0,1)$ supersymmetry
on the boundary, both not both of them. 
\section{Gauge Invariance in Presence of a Boundary } 
In the previous section we have seen that the original BLG theory is not supersymmetric  in presence of a boundary. 
We have also modified the original BLG theory by adding a boundary term such that the resultant theory preserved half of the supersymmetry even 
in presence of the boundary. In this section we will first observe that the boundary theory is not gauge invariant. 
We will then  modify this theory, by adding new boundary degrees of freedom, to make it gauge invariant even in presence of a 
boundary.

To analyse the boundary effects on the gauge invariance of the BLG theory, we first observe that the 
 matter part of the BLG theory  is gauge invariant  even in presence of a boundary, 
\begin{eqnarray}
\delta \mathcal{L}_{s, KE}  +\delta \mathcal{L}_{s, V}  &=& 
 (\delta \mathcal{L}_{KE} + \delta \mathcal{L}_{b, KE}) +  (\delta \mathcal{L}_{V}+  \delta \mathcal{L}_{b, V}) 
\nonumber \\ &=&0. 
\end{eqnarray}
However, the Chern-Simons part of the BLG theory is not gauge invariant in presence of  a boundary term.
This is because 
  the infinitesimal gauge transformation  of the Chern-Simons term is given by  
\begin{eqnarray}
  \delta \mathcal{L}_{s, CS}&=& \delta \mathcal{L}_{CS} + \delta \mathcal{L}_{b, CS}
\nonumber \\ &=&  \frac{k}{4\pi} ( \mathcal{D}_3 - \nabla^2 )[ f^{ABCD} (D^a \Lambda_{AB}\omega_a^{AB} + C^{CD,EF}_{AB}\Gamma_{CD }^a 
\Lambda_{EF} )\omega_{a CD}]_|
\nonumber \\ &=& 
  - \frac{k}{4\pi} \nabla^2 [f^{ABCD}(\nabla^a  \Lambda )_{AB}\omega_{a CD} ]_|.
\end{eqnarray}
Now this   can be written as
\begin{equation}
  \delta \mathcal{L}_{s, CS} = 
\frac{k}{4\pi} ( \mathcal{D}_3\nabla^a -(\gamma^\mu  \mathcal{D}_\mu \nabla)^a
 ) [f^{ABCD} \Lambda_{AB} \omega_{a CD} ]_|.
\end{equation}
As there is a boundary in the $x^3$ direction, we get  
\begin{eqnarray}
\delta \mathcal{L}_{s, CS}  &=& 
 \frac{k}{4\pi} (\mathcal{D}_3\nabla^a -(\gamma^\mu  \mathcal{D}_\mu \nabla)^a) [ f^{ABCD} \Lambda_{AB} \omega_{a CD}  ]_|
 \nonumber \\ &\sim
& \frac{k}{4\pi}(\mathcal{D}_3\nabla^a - (\gamma^3  \mathcal{D}_3 \nabla)^a)[ f^{ABCD} \Lambda_{AB} \omega_{a CD}]_|,
\end{eqnarray}
where $\sim$ indicates 
that we have neglected the total derivative contribution along  directions other than $x^3$, as they will not contribute.  
Thus, the gauge   transformation of this  Chern-Sioms term will be,
\begin{eqnarray}
  \delta \mathcal{L}_{s, S}
 &=&\frac{k}{4\pi}( \delta_b^a - (\gamma^3)_b^a  ){\nabla'^b} [f^{ABCD} {\Lambda' }_{AB}\omega'_{a CD}  ]_| \nonumber \\ 
  &=&   \frac{k}{2\pi}(P_- {\nabla}')^a  [ f^{ABCD} {\Lambda'}_{AB} \omega'_{a CD}  ]_|. \label{pp}
\end{eqnarray}
So, the Chern-Simons part generates a boundary term in presence of a  boundary and  the BLG theory is not gauge invariant in presence of a boundary. 

It can however be made gauge invariant by adding new degrees of freedom on the boundary. Thus,  
we define a new boundary field $v'_{AB}$  and let $v_{AB}$ its extension into the bulk. 
 We also let $v = v_{AB} T^A T^B$.  We define  generate a 
gauge transformations for $v$ as, 
\begin{equation}
  v \rightarrow vu^{-1}.
\end{equation}
We also define  $\Gamma^v$ as  the gauge transformation of $\Gamma$ generated by
$v$
\begin{equation}
 \Gamma^v_a = iv\nabla_a v^{-1}.  
\end{equation}
The  potential term for the boundary field theory can now be written as  
\begin{eqnarray}
  \mathcal{L}_{p} = \mathcal{L}_{sv , CS}
 - \mathcal{L}_{s, CS}. 
\end{eqnarray}
Here $\mathcal{L}_{sv , CS}$ is the Lagrangian obtained  by replacing $\Gamma_a$ by $ \Gamma^v_a $. 
 For $v$ close to the identity, this is a actual boundary term. However, ever for large gauge transformations this can be considered as a 
boundary term  in the sense
that in the absence of a boundary, this term will have no measurable effects.
The  Lagrangian given by the sum of this term with the ordinal supersymmetric Chern-Simons term 
$\mathcal{L}_{s, CS} + \mathcal{L}_{p}$
will now be
gauge invariant even in presence of a boundary,
\begin{equation}
 \delta \mathcal{L}_{s, CS} +  \delta \mathcal{L}_{p} =0.
\end{equation}
In fact, this potential term reduces to a generalization of the ${\cal N} = (1,0)$ Wess-Zumino-Witten model
\cite{WZW1, WZW2},  if there is no coupling of the boundary theory to the bulk fields. 
Thus,  this theory will be given by 
\begin{eqnarray}
\mathcal{L}_{p}&=&
-\frac{k}{2\pi} (P_- {\nabla}')^a C^{CD, EF}_{AB} \left[ [(v^{-1} \mathcal{D}_{+} v^{})^{AB},  
(v^{-1} \mathcal{D}_{3} v)_{CD}]\right. \nonumber \\ && \left. \times
(v^{-1} \nabla_{-a} v)_{EF} \right]_|.
\end{eqnarray}
We can now add the following supersymmetric gauge invariant kinetic term,
\begin{eqnarray}
  \mathcal{L}_{k} &=&  - \frac{k}{2\pi} 
(P_- {\nabla}')^a [ f_{ABCD}
({v'}^{-1} \nabla'_{-a} v')^{AB} ({v'}^{-1} \mathcal{D}_{+} v')^{CD} ]_|.
\end{eqnarray}
This term is gauge invariant by itself, 
$
 \delta  \mathcal{L}_{k}=0
$.
This is a generalization of the kinetic term for the $\mathcal{N} = (1,0)$ Wess-Zumino-Witten model
\cite{WZW1, WZW2}.
The Lagrangian for the boundary theory will  be  a generalization of the 
$\mathcal{N} = (1,0)$ Wess-Zumino-Witten model
\begin{equation}
  \mathcal{L}_{bt} = \mathcal{L}_{k}
 + \mathcal{L}_{p}. \label{bt}
\end{equation}
Thus, this Lagrangian also preserves the supersymmetry corresponding to $Q_-$, 
$
 \delta_{S}^-  \mathcal{L}_{bt}= 0. 
$
The Lagrangian for the final gauge and supersymmetric invariant  BLG theory  is 
\begin{equation}
\mathcal{L}_{sg, BLG} = \mathcal{L}_{s, BLG}
+ \mathcal{L}_{bt}.
\end{equation}
This Lagrangian is invariant under gauge transformations, 
\begin{equation}
 \delta  \mathcal{L}_{sg, BLG}  =0. 
\end{equation}
and preserves half of the total supersymmetry, 
\begin{eqnarray}
 \delta_{S}^-  \mathcal{L}_{sg, BLG}&=& \epsilon^{a -} Q_{- a}  \mathcal{L}_{sg, BLG} \nonumber \\ &=&0. 
\end{eqnarray}

Thus, the BLG theory is made both gauge and supersymmetric invariant in presence of a boundary by suitable coupling it to a boundary theory. 
In doing so we also obtained a generalization of the gauged  Wess-Zumino-Witten model, where the gauge symmetry is generated by a Lie $3$-algebra 
rather than an ordinary  Lie algebra.

\section{Conclusion}
In this paper we analysed the BLG theory in presence of boundaries in $\mathcal{N}=1$ superspace formalism. 
It was found that this theory is neither  invariant under the supersymmetric transformations nor invariant under the gauge transformations. 
However by coupling this theory to a boundary theory it was made both gauge and supersymmetric invariant. 
The supersymmetric and gauge variations of this boundary theory exactly canceled the boundary term generated by the 
 supersymmetric and gauge variations of this bulk theory. 

In analysing this model we developed a generalization of the gauged Wess-Zumino-Witten model, where the gauge symmetries are generated by the 
Lie $3$-algebra rather than an ordinary Lie algebra. It will be interesting to analyse certain features of this generalized gauged Wess-Zumino-Witten model
 in more detail. It is possible that this model also obeys a Kac-Moody current algebra. It will be interesting 
to analyse this algebra and study its properties. 
Furthermore, this work can be  generalise  to 
superspaces with higher amount of supersymmetry and then applied to the study of $M2$-branes with boundaries. 
As the BLG theory has also been analysed in 
$\mathcal{N } =2$  superspace \cite{214,214a}, 
and $\mathcal{N } =8$  superspace \cite{8a1, 8a2}; it would be interesting to also analyse the 
the boundary BLG theory in these superspaces.

It may be remarked that before we quantise any gauge theory we have to fix a gauge. This is done at quantum level by adding 
a gauge fixing term and a ghost term to the original classical action. This new action is invariant under a symmetry called the BRST 
symmetry. This invariance of the theory under BRST symmetry is crucial to show its unitarity.  
The BRST symmetry of Chern-Simons theory has  been thoroughly investigated
\cite{16,17} and the BRST symmetry of $\mathcal{N} = 1$   
 super-Chern-Simons theory   has also been studied in the superspace formalism \cite{18, 19}. The BRST  invariance of 
ABJM theory has also been  studied \cite{mf}. 
 It will be interesting to analyse the BRST symmetry 
for the BLG theory in presence of a boundaries in  $\mathcal{N} = 1$ superspace formalism. It is expected that the original BLG theory 
will not be invariant under the BRST transformations. However, a sum of the original BLG theory with the suitable 
boundary theory can be made  BRST invariant.

\end{document}